\documentclass[twocolumn]{elsart}
\usepackage{graphics}
\usepackage{graphicx}
\usepackage{epsfig}
\usepackage{amssymb}
\usepackage{amsmath}

\begin{document}

\begin{frontmatter}
\vspace*{-2.3mm}
\title{Measurement of the in-medium $\boldsymbol{\phi}$-meson width in proton-nucleus
collisions}\vspace*{-5mm}
\author[IKP,ITEP]{A.~Polyanskiy},
\author[IKP]{M.~Hartmann}\ead{m.hartmann@fz-juelich.de},
\author[ITEP]{Yu.T.~Kiselev}\ead{yurikis@itep.ru},
\author[RAS]{E.Ya.~Paryev},
\author[IKP]{M.~B\"uscher},
\author[IKP,TIB]{D.~Chiladze}
\author[Erlangen,Dubna]{S.~Dymov},
\author[Gatchina]{A.~Dzyuba},
\author[IKP]{R.~Gebel},
\author[IKP]{V.~Hejny},
\author[Rossendorf]{B.~K\"{a}mpfer},
\author[Basel]{I.~Keshelashvili},
\author[Gatchina]{V.~Koptev},
\author[IKP]{B.~Lorentz},
\author[Osaka]{Y.~Maeda},
\author[IKP,Dubna]{S.~Merzliakov},
\author[Gatchina]{S.~Mikirtytchiants},
\author[IKP]{M.~Nekipelov},
\author[IKP]{H.~Ohm},
\author[Rossendorf]{H.~Schade},
\author[IKP,Dubna]{V.~Serdyuk},
\author[IKP,Bonn,JLab]{A.~Sibirtsev},
\author[IKP]{H.~J.~Stein},
\author[IKP]{H.~Str\"oher},
\author[Rossendorf,MSU]{S.~Trusov},
\author[IKP,Gatchina]{Yu.~Valdau},
\author[UCL]{C.~Wilkin\corauthref{cor1}}\ead{cw@hep.ucl.ac.uk} \corauth[cor1]{Corresponding author.},
\author[ZEL]{P.~W\"ustner}
%
%
\address[IKP]{Institut f\"ur Kernphysik and J\"ulich Center for Hadron Physics,
Forschungszentrum J\"ulich, D-52425 J\"ulich, Germany}
\address[ITEP]{Institute for Theoretical and Experimental Physics,
RU-117218 Moscow, Russia}
\address[RAS]{Institute for Nuclear Research, Russian Academy of
Sciences, RU-117312 Moscow, Russia}
\address[TIB]{High Energy Physics Institute, Tbilisi State University,
0186 Tbilisi, Georgia}
\address[Erlangen]{Physikalisches Institut II, Universit{\"a}t
Erlangen-N{\"u}rnberg, D-91058 Erlangen, Germany }
\address[Dubna]{Laboratory of Nuclear Problems, Joint Institute for
Nuclear Research, RU-141980 Dubna, Russia}
\address[Gatchina]{High Energy Physics Department, Petersburg Nuclear
Physics Institute, RU-188350 Gatchina, Russia}
\address[Rossendorf]{Forschungszentrum Dresden Rossendorf, D-01314 Dresden, Germany}
\address[MSU]{Skobeltsyn Institute of Nuclear Physics, Lomonosov Moscow
State University, RU-119991 Moscow, Russia}
\address[Basel]{Department of Physics, University of Basel,
Klingelbergstrasse 82, CH-4056 Basel, Switzerland}
\address[Osaka]{Research Center for Nuclear Physics, Osaka
University, Ibaraki, Osaka 567-0047, Japan}
\address[Bonn]{Helmholtz-Institut f\"ur Strahlen- und Kernphysik
and Bethe Centre for Theoretical Physics, Universit\"at Bonn, D-53115
Bonn, Germany}
\address[JLab]{Excited Baryon Analysis Center (EBAC), Thomas Jefferson National
Accelerator Facility, Newport News, Virginia 23606, USA}
\address[UCL]{Physics and Astronomy Dept., UCL,
Gower Street, London WC1E 6BT, U.K.}
\address[ZEL]{Zentralinstitut f\"ur Elektronik,
Forschungszentrum J\"ulich, D-52425 J\"ulich, Germany}
%
%
\vspace*{1cm}
\begin{abstract}
The production of $\phi$ mesons in the collisions of 2.83~GeV protons
with C, Cu, Ag, and Au at forward angles has been measured via the
$\phi\to K^+K^-$ decay using the COSY-ANKE magnetic spectrometer. The
$\phi$ meson production cross section follows a target mass
dependence of $A^{0.56\pm 0.02}$ in the momentum region of
0.6--1.6~GeV/c. The comparison of the data with model calculations
suggests that the in-medium $\phi$ width is about an order of
magnitude larger than its free value.
\end{abstract}
\end{frontmatter}


How the properties of light vector mesons change when they are in a
strongly interacting environment has been a very active research
field for several years~\cite{Leupold:2009kz,Hayano:2008vn},
especially in connection with the question of the partial restoration
of chiral symmetry in hot/dense nuclear matter.

The most interesting case is that of the $\phi(1020)$ meson, whose
width in vacuum of 4.3~MeV/$c^2$ is narrow compared to those of other
nearby resonances. Small modifications in medium should therefore be
experimentally observable. Hadronic
models~\cite{Oset:2000eg,Cabrera:2002hc,Klingl:1997tm} predict an
increase in the width of low-momentum $\phi$ mesons in cold nuclear
matter at nuclear saturation density by up to a factor of ten
compared to the free value, whereas an insignificant mass shift is
expected in both these models and in QCD sum rule
studies~\cite{Hatsuda:1991ez,Zschocke:2002mn}.

Dileptons from $\phi \to e^{+}e^{-}/\mu^{+} \mu^{-}$ decays experience
no strong final-state
interactions in a nucleus so that any broadening of the $\phi$ spectral
shape could be directly tested by
examining dilepton mass spectra produced by elementary
($\gamma,\,\pi,\, p$) probes, provided the necessary cuts are applied
on the low $\phi$ momenta~\cite{Paryev:2005wh,Muto:2005za}. These
reactions are less complicated than heavy-ion collisions
because they proceed in cold static matter of a well-defined density.
Furthermore, it has been
argued that the sensitivity of such reactions to in-medium changes of
hadron properties should be comparable to those of nucleus-nucleus
collisions~\cite{Mosel:1998hk}.

Measurements of dilepton invariant mass distributions are, however,
difficult due to the low branching ratios for leptonic
decays. The KEK-PS-E325
collaboration measured $e^{+}e^{-}$ invariant mass spectra in the
$\phi$ region in proton-induced reactions on carbon and copper at
12~GeV and deduced a mass shift of 3.4\% and a width increase
by a factor of 3.6 at normal nuclear matter density for $\phi$
momenta around 1~GeV/$c$~\cite{Muto:2005za}.

An alternative method of studying the in-medium broadening of the
$\phi$ meson has been considered both
theoretically~\cite{Barz:2003dd,Magas:2004eb,Cabrera:2003wb,Muhlich:2005kf,Sibirtsev:2006yk,Paryev:2008ck}
and experimentally~\cite{Ishikawa:2004id,Wood:2010ei}. The variation
of the $\phi$ production cross section with atomic number $A$ depends
on the attenuation of the $\phi$ flux in a nuclear target which, in
turn, is governed by the imaginary part of the $\phi$ in-medium
self-energy or width. In the low-density approximation, this width
can be related to the $\phi N$ total cross section. The advantage of
this method is that one can exploit the large $K^{+}K^{-}$ branching
ratio $(\approx 50\%)$ in order to identify the $\phi$ meson in
production experiments on nuclear targets. Owing to the small energy
release in this channel, any renormalization of the $\phi$ is very
sensitive to the in-medium modification of kaons and antikaons, a
subject which is also of great current interest.

An unexpectedly large in-medium $\phi N$ total cross section of about
35~mb was inferred from measurements of $K^+K^-$
pairs photoproduced on Li, C, Al and Cu targets at
SPring8~\cite{Ishikawa:2004id}. In the low-density approximation,
this implies a larger in-medium
$\phi$ width than the KEK result~\cite{Muto:2005za}.
Very recent data on $\phi$ photoproduction at JLab~\cite{Wood:2010ei} also indicate
a substantial in-medium broadening, though the precision in both cases is not
sufficient to rule out the KEK result~\cite{Muto:2005za}.

In an attempt to clarify the situation, measurements have been carried out at the
ANKE-COSY facility~\cite{Barsov:2001xj,Hartmann:2007ks} with a proton
beam, detecting the $\phi$ through its $K^{+}K^{-}$ decay. The main
goal of these measurements was to obtain values of the so-called
transparency ratio,
\begin{equation}
\label{eq::eq1}%
R = \frac{ 12~\sigma_{pA \to \phi X'}}{A~\sigma_{p{\rm C} \to \phi
X}}\,,
\end{equation}
normalized to carbon. Here $\sigma_{pA \to \phi X'}$ and
$\sigma_{p{\rm C} \to \phi X}$ are inclusive cross sections for
$\phi$ production in $pA$ and $p$C collisions, respectively.

The experiment was performed with a $2.83~$GeV proton beam, which is
only a little above the $\phi$ production threshold in
proton-nucleon collisions. The contributions from channels where
there is an additional pion produced are therefore expected to be small.
The targets used were light and medium nuclei C and
Cu, as in the KEK~\cite{Muto:2005za} and
SPring8~\cite{Ishikawa:2004id} measurements, but also the heavier
Ag and Au, where distortion effects should be much stronger.
The very narrow targets had thicknesses around 10--30~$\mu$m.

The COSY-ANKE spectrometer, located at an internal target position of
COSY, is composed of three dipole magnets. $D1$ and $D3$ bend the
direct proton beam from the undisturbed COSY orbit to the ANKE target
and return it back, respectively. The analyzing magnet $D2$, placed
between $D1$ and $D3$, deflects reaction products into the detection
systems placed to the right and left of the beam to register,
respectively, positively and negatively charged ejectiles, in this
case $K^+$ and $K^-$. Positive kaons were first selected using a
dedicated detection system that can pick out a $K^+$ against a
$\pi^{+\!}$/$p$ background that is $10^5$ more
intense~\cite{Buescher:2002zc}. The coincident $K^{-}$ was
subsequently identified from the time-of-flight difference between
the stop counters in the negative and positive detector systems,
these selections being carried out within $\pm 3~\sigma$
bands~\cite{Hartmann:2007ks}.

\begin{figure}[t]
  \begin{center}
{\includegraphics[width=0.9\columnwidth, clip]{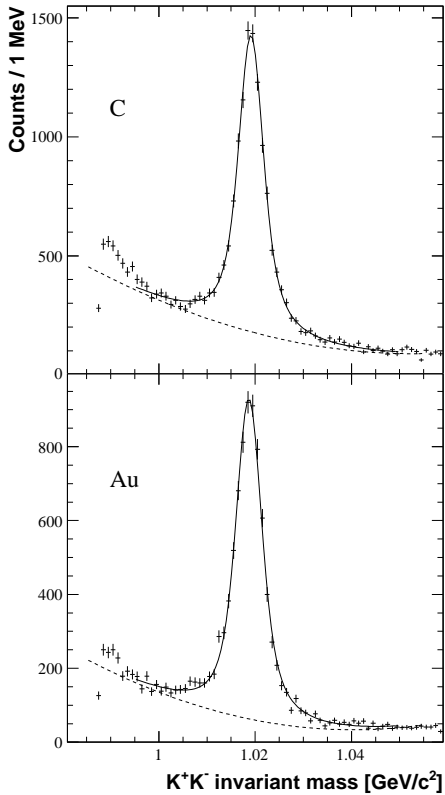}}
\vspace*{-2mm}%
\caption{Invariant mass distributions for $K^{+}K^{-}$ pairs produced
in $p$C and $p$Au collisions. The experimental data are not
acceptance-corrected. The dashed lines are second-order polynomial
representations of the backgrounds in the region of the $\phi$
peak.}\label{fig:KKIM}
\end{center}
\end{figure}

The $K^{+}K^{-}$ invariant mass spectra for the $pA\to K^+K^- X$
reaction look similar for the four targets and the results for the C
and Au targets are presented in Fig.~\ref{fig:KKIM}. In all cases
there is a clear $\phi$ peak sitting on a background of non-resonant
$K^+K^-$ production together with a relatively small number of
misidentified events.

The ANKE spectrometer only registers $\phi$ mesons at small
laboratory polar angles, $0^{\circ} < \theta_{\phi} < 8^{\circ}$,
over the limited momentum range, $0.6~\textrm{GeV}/c < p_{\phi} <
1.6~\textrm{GeV}/c$. To study the $A$-dependence of the transparency
ratio, the numbers of $\phi$ events that fall within this acceptance
window were first evaluated for every target. For this purpose, each
mass spectrum was fitted by the sum of a Breit-Wigner function with
the natural $\phi$ width, convoluted with a Gaussian resolution
function with $\sigma=1~\textrm{MeV}/c^2$, and a background function.
This procedure therefore concentrates on the vast majority of the
$\phi$ mesons that decay outside of the nucleus. Typical examples of
the resulting fits are shown in Fig.~\ref{fig:KKIM}.

The systematic uncertainties were studied by varying the fit region,
binning, and mass scale, and changing the background curve from
linear, quadratic to cubic. Their values for each ratio were then
averaged and the final systematic uncertainty taken to be equal to $3
\sigma$. The number of reconstructed $\phi$ mesons for each target
was between 7000 and 10000.

The relative luminosity for each target was derived by measuring
simultaneously the fluxes of $\pi^+$ mesons with momenta between 475
and 525~MeV/$c$ in the angular cone
$\theta_{\pi}<4^\circ$. Since the double-differential cross section
for $\pi^+$ production has not been measured at 2.83~GeV, we
parametrized the available data~\cite{pions}
at seven proton energies in the range 1--5.6~GeV in the form
\begin{equation}%
\label{eq::eq2}%
\sigma_{A} =\sigma_{0}A^{\alpha}.%
\end{equation}%
The interpolation of these fits to 2.83~GeV yielded an exponent
$\alpha_{\pi}$=$0.38 \pm 0.02$, which allowed us to normalize
the ratios of the numbers of measured $\phi$ mesons. Since the
acceptance corrections in ANKE are essentially target-independent,
this corresponds to the ratio of the cross sections for $\phi$
production in $pA$ and $p$C collisions in the ANKE acceptance window.
The resulting transparency ratios given in Table~\ref{tab::table1}
correspond to production rates that
follow the power law of Eq.~\eqref{eq::eq2} with $\alpha_{\phi} =
0.56\pm0.02$.

\begin{table}[ht]
\caption{The measured transparency ratio $R$ of Eq.~\eqref{eq::eq1}
in the acceptance window of the ANKE spectrometer. The first errors
are statistical and the second systematic. The latter arise mainly
from the evaluation of the numbers of $\phi$ events and the relative
normalizations. \label{tab::table1}} \vspace{2mm}
\begin{center}
\begin{tabular}{|c|c|}
\hline%
$A$/C & R\\
\hline
Cu/C & $0.479\pm 0.011\pm0.035$\\
Ag/C & $0.387\pm 0.009\pm0.033$\\
Au/C & $0.292\pm 0.007\pm0.021$\\
\hline
\end{tabular}
\end{center}
\end{table}

\begin{figure}[t]
  \begin{center}
{\includegraphics[width=0.9\columnwidth, clip]{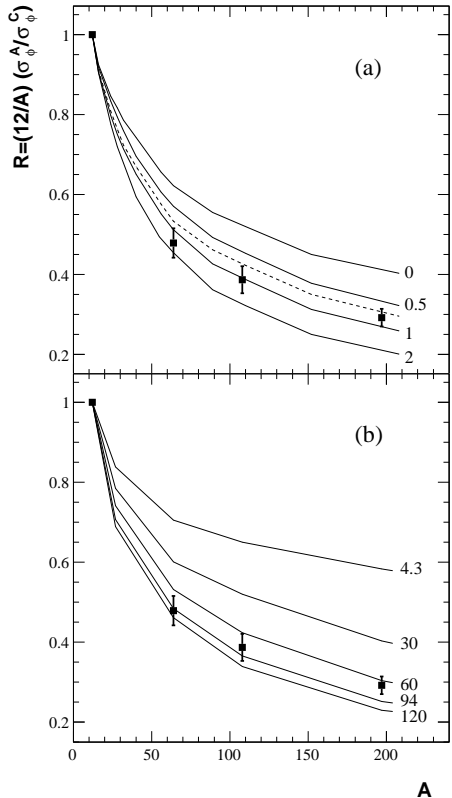}}
\vspace*{-2mm}%
\caption{Comparison of the measured transparency ratio $R$ as a
function of atomic number $A$ with (a) the Valencia
calculations~\cite{Magas:2004eb}, where the input width of
24~MeV/$c^2$ was multiplied by factors 0, 0.5, 1, and 2, as
indicated. (b) Predictions of the Paryev model~\cite{Paryev:2008ck}
for different $\phi$ widths. For the rest of the notations, see the
text. } \label{fig:adep}
  \end{center}
\end{figure}

Any interpretation of the transparency ratio has to rely on a
detailed theoretical treatment. In Fig.~\ref{fig:adep}a we compare
the experimental data with the results of calculations performed by
the Valencia group within the local Fermi sea approach, using the
eikonal approximation to account for the absorption of the outgoing
$\phi$ meson~\cite{Magas:2004eb}. The calculations have been done for
$N=Z$ nuclei in a single-step ($pN \to pN\phi$) model, using
predictions of the group~\cite{Cabrera:2002hc,Cabrera:2003wb} for the
imaginary part of the $\phi$ self-energy in nuclear matter. This
corresponds to a total width of $\Gamma=28$~MeV/$c^2$ for a $\phi$ at
rest at saturation density $\rho_{0}$. Figure~\ref{fig:adep}a shows
the predictions when this width, without the contribution from the
free $\phi$ decay, is multiplied by factors of 0, 0.5, 1 and 2. It
should be noted that these estimates were carried out over the whole
available phase space. A fit to our data within this model yields the
value of $\Gamma={27}^{+5}_{-3}$~MeV/$c^2$.

The contributions to $\phi$ production from two-step processes, with
nucleon and $\Delta$ intermediate states, have been estimated for the
nominal width conditions~\cite{Magas:2004eb} and these lead to an
increase in the transparency ratio $R$, which is shown by the dashed
line in Fig.~\ref{fig:adep}a. If we take into account the isospin
corrections for the $N \ne Z$ nuclei according to the prescription
from Ref.~\cite{Magas:2004eb}, then a fit to our data within the
extended model leads to $\Gamma={45}^{+17}_{-9}$~MeV/$c^2$. This
about 50\% greater than that predicted in~\cite{Cabrera:2002hc}, but
is in good agreement with the theoretical estimates of Klingl
\textit{et al.}~\cite{Klingl:1997tm}, who suggest $\Gamma\approx
45$~MeV/$c^2$.

In an alternative theoretical approach, Paryev~\cite{Paryev:2008ck}
analyzed $\phi$ production in proton-nucleus reactions by considering
primary proton-nucleon ($pp\to pp\phi$, $pn\to pn\phi$, $pn\to
d\phi$) and secondary pion-nucleon ($\pi N \to \phi N$) processes in
the framework of a nuclear spectral function model. This calculation
takes into account the Fermi momentum of the struck target nucleon
and the removal energy distribution. It uses the new measurements of
the $pp\to pp\phi$ and $pn\to d\phi$
reactions~\cite{Maeda:2008,Maeda:2006} and estimates of the cross
section difference between $pn\to pn\phi$ and $pp\to
pp\phi$~\cite{Kaptari:2005}. The total in-medium width of the $\phi$
meson in its rest frame was assumed to be momentum-independent, as
suggested by the Valencia calculations~\cite{Cabrera:2003wb}. Results
have been obtained for different values of this width, as indicated
by the curves in Fig.~\ref{fig:adep}b. It should be emphasized that
the ANKE kinematical cuts on the laboratory $\phi$ momenta and
production angles were included.

Fitting the data with the full Paryev model~\cite{Paryev:2008ck},
yields a value of ${73}^{+14}_{-10}$~MeV/$c^2$ for the in-medium
width of a moving $\phi$ meson in its rest frame at
$\rho_{0}=0.16$~fm$^{-3}$. This corresponds to $\approx 50$~MeV/$c^2$
in the nuclear rest frame when the $\phi$ has a momentum of
1.1~GeV/$c$, which is typical for the ANKE conditions. For a $\phi$
at rest in this frame at nuclear density $\rho_0$ this gives
$\Gamma\approx 73$~MeV/$c^2$, in line with the original model
assumptions~\cite{Paryev:2008ck}.

In order to elucidate the discrepancies between the two models, a
full phase space calculation within the approach of
Ref.~\cite{Paryev:2008ck} has been repeated for $N=Z$ nuclei, keeping
only the primary $\phi$ production processes $pN \to pN \phi$. With a
$\phi$ total width of 30~MeV/$c^2$, the results are close to those
obtained in~\cite{Magas:2004eb} under similar conditions and
represented by the line marked `1' in Fig.~\ref{fig:adep}a. Moreover,
it is found that the effects arising from the cuts imposed by the
ANKE acceptance window are relatively minor. Hence the differences
between the two theoretical approaches must be ascribed to the
effects of secondary processes, which can have quite different $A$
dependences~\cite{Magas:2004eb,Paryev:2008ck,Sibirtsev:2008ib}. The
first results from the ongoing Rossendorf Boltzmann-Uehling-Uhlenbeck
(BUU) calculations~\cite{Schade:2010} lie between those of the two
models. It is therefore clear that, in order to pin down $\phi$
medium effects better, it is necessary to improve our understanding
of the $\phi$ production mechanisms in nuclei.

We can compare our results with those obtained in the SPring8
experiment~\cite{Ishikawa:2004id}. The analysis of these data within
the Giessen BUU transport model shows a strong influence of the
nuclear environment on the properties of the
$\phi$~\cite{Muhlich:2005kf}. In the low-density approximation, the
$\phi N$ total cross section of ${35}^{+17}_{-11}$~mb extracted from
these data corresponds to an in-medium width of about 80~MeV/$c^2$
for the ANKE conditions. The width of 50~MeV/$c^2$ deduced from our
data on the basis of the approach of Ref.~\cite{Paryev:2008ck} is not
inconsistent with this, taking into account the uncertainties in both
results (see~\cite{Oset:2005ag}). It is also in line with that
deduced from the JLab measurements~\cite{Wood:2010ei} but clearly
larger than that found in the KEK experiment~\cite{Muto:2005za}.

The direct fitting of the full models to the measured transparency
ratios yields values of the total $\phi$ in-medium width of about
45~MeV/$c^2$~\cite{Magas:2004eb} and
73~MeV/$c^2$~\cite{Paryev:2008ck}, with the BUU
calculations~\cite{Schade:2010} falling between. In order to draw
conclusions whether the observed $\phi$ width exhibits non-trivial
medium effects, one has to compare this with naive expectations.
Starting from a free $\phi N$ cross section of $\approx 10$~mb, the
authors of Ref.~\cite{Muhlich:2005kf} found, for vanishing $\phi$
momenta in the low-density approximation, a collision width of
$\approx 18$~MeV/$c^2$ at density $\rho_{0}$. Taking into account the
free $\phi$ width of 4.3~MeV/$c^2$, this leads to a total width of
$\approx 22$~MeV/$c^2$. Independent of the model used for the
analysis, our values clearly exceed this and leave significant space
for non-trivial effects arising from the nuclear medium. Effects from
$\phi/\omega$ mixing~\cite{Sibirtsev:2006yk}, which have been invoked
to explain non-spectator events in the $\gamma d\to \phi pn$
reaction~\cite{Qian:2009vi}, can be even more important in our case
because the hadronic production of $\phi$ mesons is suppressed by the
OZI rule~\cite{OZI}. It should, however, be stressed that the
transparency ratio measurements do not allow one to disentangle the
various mechanisms for medium modification.

In summary, we have performed a high statistics measurement of the
transparency ratio for $\phi$ meson production with 2.83~GeV protons
on C, Cu, Ag and Au targets. The production cross section was found
to vary like $A^{0.56\pm 0.02}$ for $\phi$ in the momentum range
0.6--1.6~GeV/$c$. Values of the $\phi$ width in nuclear matter were
obtained by comparing the data with the two available
models~\cite{Magas:2004eb,Paryev:2008ck}. The results found indicate
a substantial increase in the total $\phi$ width in the nuclear
environment. Further theoretical and experimental efforts are needed
to reach a better understanding of the phenomenon of $\phi$
renormalization in nuclear matter.

Support from A.~Wirzba and other members of the ANKE Collaboration,
as well as the COSY machine crew, are gratefully acknowledged. The
calculations performed for us by the Valencia group have been very
helpful in the interpretation of our results. This work has been
partially financed by the BMBF, COSY FFE, DFG, and RFBR.

\vspace*{-5mm}

\end{document}